\documentclass[10pt,journal,compsoc]{IEEEtran}
\usepackage{amsmath}
\usepackage{subfigure}
\usepackage{multirow}
\usepackage{graphicx}
\usepackage{algorithm}
\usepackage{algpseudocode}

\ifCLASSOPTIONcompsoc
  \usepackage[nocompress]{cite}
\else
  \usepackage{cite}
\fi
\ifCLASSINFOpdf
\else
\fi
\begin{document}
%
\title{DeepSpace: An Online Deep Learning Framework for Mobile Big Data to Understand Human Mobility Patterns}
%
%
%
%

\author{Xi~Ouyang\IEEEauthorrefmark{1},
        Chaoyun~Zhang\IEEEauthorrefmark{1},
        Pan~Zhou\IEEEauthorrefmark{2},~\IEEEmembership{Member,~IEEE},
        Hao~Jiang,~\IEEEmembership{Member,~IEEE},
        and~Shimin~Gong,~\IEEEmembership{Member,~IEEE}
\IEEEcompsocitemizethanks{
\IEEEcompsocthanksitem Xi Ouyang and Pan Zhou are with the School of Electronic Information and Communications, Huazhong University of Science and Technology, Wuhan 430074, China. Pan Zhou\IEEEauthorrefmark{2}, is the corresponding author of this paper. Email: ouyang@hust.edu.cn, panzhou@hust.edu.cn.
\IEEEcompsocthanksitem Chaoyun Zhang\IEEEauthorrefmark{1}, co-author of this paper, is with School of Informatics, The University of Edinburgh, Edinburgh, UK. Email: chaoyun.zhang@ed.ac.uk.
\IEEEcompsocthanksitem Hao Jiang is with the School of Electronics Information, Wuhan University, Wuhan 430074, China. E-mail: jh@whu.edu.cn.
\IEEEcompsocthanksitem Shimin Gong is with Shenzhen Institutes of Advanced Technology, Chinese Academy of Sciences, Shenzhen, Guangdong, China. E-mail: sm.gong@siat.ac.cn.}
}
\IEEEtitleabstractindextext{%
\begin{abstract}
In the recent years, the rapid spread of mobile device has create the vast amount of mobile data. However, some shallow-structure models such as support vector machine (SVM) have difficulty dealing with high dimensional data with the development of mobile network. In this paper, we analyze mobile data to predict human trajectories in order to understand human mobility pattern via a deep-structure model called ``DeepSpace". To the best of out knowledge, it is the first time that the deep learning approach is applied to predicting human trajectories. Furthermore, we develop the vanilla convolutional neural network (CNN) to be an online learning system, which can deal with the continuous mobile data stream. In general, ``DeepSpace" consists of two different prediction models corresponding to different scales in space (the coarse prediction model and fine prediction models). This two models constitute a hierarchical structure, which enable the whole architecture to be run in parallel. Finally, we test our model based on the data usage detail records (UDRs) from the mobile cellular network in a city of southeastern China, instead of the call detail records (CDRs) which are widely used by others as usual. The experiment results show that ``DeepSpace" is promising in human trajectories prediction.
\end{abstract}

}

\maketitle

\IEEEdisplaynontitleabstractindextext

%
\IEEEpeerreviewmaketitle

\ifCLASSOPTIONcompsoc
\IEEEraisesectionheading{\section{Introduction}\label{sec:introduction}}
\else
\section{Introduction}
\label{sec:introduction}
\fi

%
%
%
%
\IEEEPARstart{O}{ver} the course of the past decade, mobile big data has a explosive growth due to the fast development of various mobile devices (e.g., iPhone, Android phone, Google Glass, iWatch etc.). Global mobile data traffic grew by 69\% in 2014, according to Cisco Visual Networking Index (VNI) \cite{cisco}. Meanwhile, the mobile data traffic created in last years was nearly 30 times bigger in terms of size than the entire global Internet in 2000. It is amazing that global mobile data traffic will increase nearly tenfold between 2014 and 2019 based on the Cisco VNI's forecast. At the same time, with the rapid development of the Internet-of-Things (IoT), much more data is automatically generated by millions of machine nodes with growing mobility, such as sensors carried by moving objects or vehicles \cite{cisco1}. As we shall see, mobile big data has already penetrated to every aspects in our lives. More and more people conduct their social activities, watch videos, buy things via their mobile devices. Mobile big data is a gold mine containing large amounts of individual information \cite{information1,information2,information3}.

Mobile data is naturally generated by various mobile devices carried by human, which means that it involves the mobility patterns of each carrier. Understanding human mobility pattern is valuable in many geographical applications, such as transportation planning and epidemic disease controlling \cite{useful1}. For example, Uber drivers can reduce waiting time for passengers if the company can provide the precise trajectory
information of customers. A research team finds a 93\% potential predictability for each individual mobility across their whole research users \cite{nature}. This conclusion is based on the research from Chaoming Song and Albert-L\'{a}szl\'{o} Barab\'{a}si who study the mobility patterns of many anonymous mobile phone users. Their experiments show that there are few freewheeling style men in the crowd. Although people may have significant differences in the moving patterns, their moving patterns are predictable for each individual. Besides, this prediction is largely independent of the distance human cover on a regular basis. Based on the unique characteristics of human moving behaviour, we propose an effective deep-learning based model to predict the human trajectories.

Recently, deep learning has achieves marvelous results in many field (e.g., computer vision, speech recognition, etc.) \cite{dlnew1,dlnew2,dlnew3}. It shows its distinct advantages for the complex and massive data with the benefit of a deep nonlinear network structure. Mobile data has complex spatiotemporal relation and it is usually high-dimensional. The shallow-structure models like support vector machine (SVM) have difficulty dealing with high dimensional data with the development of mobile network, because shallow-structure algorithms need large amount of priori knowledge to extract feature manually. In contrast, deep learning just like the "black magic": it has been shown to outperform state-of-the-art machine learning algorithm for such complex data. Considering the power of deep learning, we develop our model using a convolutional neural network (CNN) which is one of the popular and effective models in deep learning. There are two key advantages in applying CNN to human mobility pattern task:
\begin{itemize}
\item Local Dependency: CNN is able to exploit local dependencies of nodes in the human trajectories. In the field of computer vision, CNN can capture the local dependency of visual information. There has strong correlation between the nearby pixels in images. The nearby positions in the human trajectories are also likely to be relative. In a specific time interval, a human can only move to the specific positions based on the past positions rather than the random positions because of the limited mobility and time.
\item Scale Invariance: CNN ensures feature scale to be invariant. In image recognition, the training images might have different scales. When predicting human mobility pattern, a person can have similar trajectories every day, but the time interval between positions and staying time in each position is often different. CNN can preserve spatial feature scale to be invariant in different temporal paces.
\end{itemize}

\begin{figure*}[!htb]
\centering
\includegraphics[scale=.4]{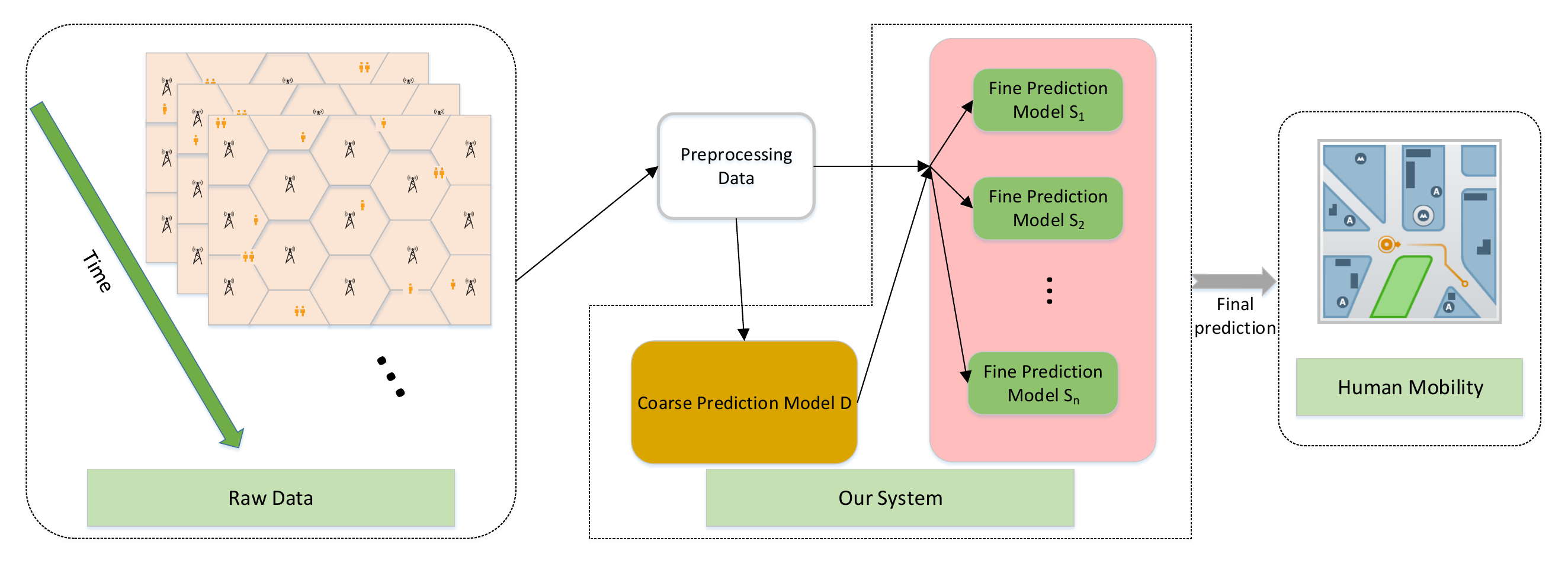}
\caption{The framework of our system. It is a hierarchical framework (consist of the coarse model and fine models). This system suits the characteristics of spatial mobile data and works as an online learning system.}
\label{fig:digraph1}
\vspace{-.8em}
\end{figure*}

Furthermore, this model we design is an online learning structure for the continuous mobile big data stream. We call our model ``DeepSpace", and the overall structure is shown in Fig. \ref{fig:digraph1}. Our model is a hierarchical framework, and the model completes the final predicting via two sub-architectures (the coarse prediction model and fine prediction models). We will describe this system in the following paper in detail to explain how this system works as an online learning system and suits the spatial characteristics of the mobile data. The contribution of this work are summarized as follows:
\begin{itemize}
\item ``DeepSpace" is an online learning model, which can deal with the continuous mobile data stream in real time. Meanwhile, we can run this model in parallel, which is efficient even only limited computing resource is avaliable.
\item To the best of our knowledge, ``DeepSpace" is the first deep learning system that is redesigned to adjust to the spatial characteristics of the mobile data.
\end{itemize}

The remainder of the paper is organized as follows. In Section 2, we first introduce related work with respective to human mobility prediction, deep learning and online learning. Meanwhile, we discuss the motivation in Section 2. Next, Section 3 devotes to the proposal of ``DeepSpace" model and its inference process. In Section 4, we describe our algorithm in detail. The performance of ''DeepSpace" is evaluated through experiments on our mobile dataset, and compared with the vanilla CNN in Section 5. Finally, we will give a brief conclusion and summarize the future in the last section.

\section{Related work and motivation}
\subsection{Related work}
\textbf{Predicting human mobility.} Understanding and predicting human mobility catches the attention and interests of researchers and practitioners \cite{mobility1,mobility2,mobility3,mobility4}. Ahas et al. find that there is significance difference in intensity of activities for each person, but everyone is regular within 24 hours \cite{timemove}. Moreover, the research from Song et al. shows that personal mobility is non-random, which indicates that people always appear in several specific places. In spite of the diversity of behavior, human mobility always obeys the simple reproduction mode and a high degree of predictability \cite{nature,song}. Sevtsuk uses the erlang from base stations as indicators to measure the population distribution, and reach a conclusion that human activities have the universal law over different time period (e.g., hour, day, week, etc.) \cite{sevt}. In a nutshell, human' spatial mobility is proved to have the following special features: 1) Regional. The probability of leaving a small area decreases with time. 2) Strongly regular and predictable. There are several frequent areas for individuals.

\textbf{Deep learning.} Recently, deep learning make many breakthroughs in different research fields, and has drawn great attention from both industrial and academic circles \cite{dlnature}. It is proved to be very powerful in both supervised and unsupervised learning fields \cite{deep1,dloverview}. In fact, the artificial neural network, which is the predecessor of deep learning, can be traced back to the new perceptron proposed by Fukushima Kunihiko in 1980 \cite{ann}. In 1989, Yann et al. \cite{mnist} use a convolutional neural network to recognize the handwritten numeral. Although their model obtains a great accuracy in handwritten numerals dataset, it requites too many computation and memory resources, which can run for three days in the 1990s. However, after 2006, due to the excellent development
of neural network \cite{dl1,dl2,dl3}, the theory of deep-structure models have major breakthroughs and solve the vanishing gradient problem \cite{grad}. Meanwhile, accompanied by the development of hardware, the applications of deep learning boom since 2006. Especially in the field of computer vision, Hinton and his students \cite{imagenet} greatly reduce the top-1 and top-5 error rates of ImageNet via many efficient improvements in CNN. Applying CNN to non-image classification has also made rapid progress. In \cite{cnnhuman,deep2,deep3}, Ming et al. propose an approach based on CNN to recognize human activities (e.g., ``jogging", ``walking", ``ascending stairs", etc.). Their CNN-based approach is practical and achieves higher accuracy than existing state-of-art methods.

\textbf{Online learning.} There are three representative papers \cite{online1,online2,online3} in transforming deep learning into a online learning model. In \cite{online1}, Zhou et al. use the denoising autoencoder \cite{dae} as a building block for online learning. This algorithm is effective in recognizing new patterns via adding new feature mappings to the existing feature set and merging parts of the existing feature which are redundant. However, fast speed data flow can not be solved in a timely manner because the model has too many parameters and it can not run in parallel. In many online learning systems, there is a key problem called ``catastrophic forgetting" \cite{online3} which refers to the model "forgets" how to perform the old task after being trained with data of new classes. This is a problem that calls for immediate solution when we design an online deep learning algorithm. To address this problem, a hierarchical training algorithm for CNN is proposed by Xiao et al \cite{online2}. When the new data comes, they will be used for training in the coarse-grained superclasses firstly, and then those return final prediction within a superclass. This model can run in parallel which is especially suitable for the massive data stream.

\subsection{Motivation}
The aim of our mobile big data analysis is to answer this question: where do people go next? \cite{mobile}. Predicting human mobility is a significant question both in academia and industry as mentioned above. This part will focus on the advantages of developing an online learning framework compared with an offline framework. An online learning system is more practical when considering some real scenes. For instance, when predicting the human trajectories using the mobile data of base stations in real time, we may find too many people gather around in a small area in a time interval. This may exceed the maximum load capacity of the base station in this small area. Then, according to the real-time warning of our system, the mobile operators can send an emergency repair team to that area in advance to prevent the base station from breaking down. The system we develop can make response in time if anything goes wrong, which is significant in daily life.

We will describe our motivations of applying deep learning to build our system in this paragraph. Deep learning has been proved to have huge advantages in three aspects: 1) Dealing with massive and high-dimensional data. 2) Processing the heterogeneous and multiclass data. 3) Deep learning can extract features automatically without the complex feature selection. Number three is the biggest motivation to drive us use a deep learning framework instead of those shallow machine learning algorithms (e.g., logistic regression, SVM, Naive Bayes). Deep learning can extra human mobility features from mobile data without any domain knowledge, which indicates that we can ignore the possible interaction between sociology and human mobility. Then we can focus on the predicting task itself. Meanwhile, we can develop an end-to-end structure because it is not necessary to extra features manually at first.

In this paper, we choose the CNN to build our system ``DeepSpace". We will discuss about it in detail in the next section.
\section{DeepSpace model}
In this section, we will introduce our proposal of ``DeepSpace" model and its inference process. In this article, we present a new solution to predict human's moving path via a deep learning model. Although the remarkable achievements of deep learning indeed show its tremendous potential in application for mobile data, there is still the great bulk of work to adapt deep learning to meet the characteristics of mobile data. In this paper, we develop an online CNN model that can process the online continuous mobile data flow. Meanwhile, our algorithm is in accordance with the spatial feature of the mobile data perfectly. As such, we call this online CNN algorithm ``DeepSpace".

Inspired by \cite{online1} and \cite{online2}, we build this online model to deal with the continuous mobile data in two different ways:
\begin{itemize}
\item Finding an optimal feature set size for the online data. Specifically, this model is able to add new features to minimize the objective function's residual and merge similar features to obtain a compact feature representation in order to prevent over-fitting.
\item Changing the CNN to a higher level hierarchical structure. This algorithm divides an overall CNN into several sub-models: the low-layer model provides the input for high-layer models, high-layer models completes the final predictions.
\end{itemize}

In the mobile network, we usually deal with the continuous data stream from base stations rather than in the sever clusters. For predicting human trajectories, the real-time character is a significant factor that we have to consider, which means that we should make predictions promptly subject to the limited computing time and power. When we develop a hierarchical CNN, all component models can be trained in parallel, which can speed up the learning process. Moreover, it will be hard to train the model with restrictions of computation and memory resources if using just a single CNN. The hierarchical model balances the time of training and base station's computer power. Besides, thanks to parallel structure, this model can significantly reduce the dimension of data. We make predictions based on the data gathered from base station.

Another important motivation of building a hierarchical CNN is that hierarchical model is more suitable for characteristic of this spatial data. First, the hierarchical architecture is mapped to the structure of spatial data administrative division (like New York City - Manhattan). In specific, the high-layer models of ``DeepSpace" refer to the senior administrative region. Similarly, the low-layer models refer to the junior administrative region. Secondly, the process of training this hierarchical architecture mirrors the hierarchical and iterative inference process that human brain to be adept at, as we do a crude prediction first, and then refine it further. For instance, we can not identify the exact position at which the man will arrive at the next moment. Intuitionally, we might locate the man at New York' Manhattan firstly, then we confirm the man's exact position just at the area of Manhattan. Based on \cite{DeepID}, it will contribute to the performance of predicting via dividing data into area patches. Because of the distinctive characteristics of mobile big data, ``DeepSpace" is not only an online CNN architecture, but also a suitable structure for the spatial data. We will discuss the overall system in the next section.
\section{System design}
In the following section, we will describe our model in detail.

\subsection{System overview}
\begin{figure}[!htb]
\centering
\includegraphics[height=7.5cm, width=7cm]{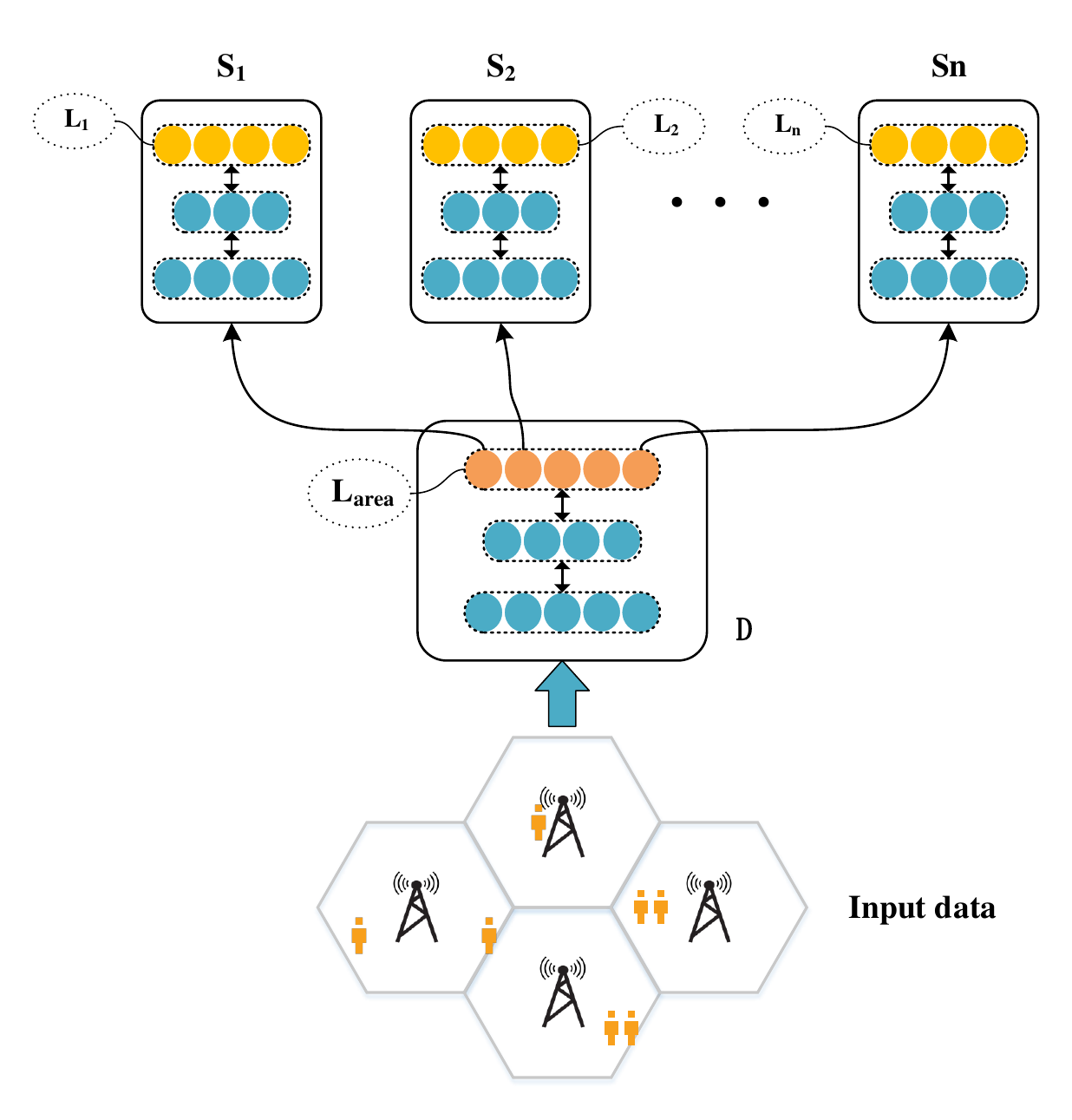}
\caption{``DeepSpace" model. The component model $D$ is the coarse prediction model: doing the crude predicting and leading the data flow. The component models \{$S_1$, $S_2$, ..., $S_n$\} are the fine prediction models: finishing the final prediction.}
\label{fig:digraph2}
\vspace{-.8em}
\end{figure}

The overall structure of ``DeepSpace" model is shown in Fig. \ref{fig:digraph2}.

We separate the whole network into two layers sub-models. The lower-layer sub-model $D$ closed to the input data layers is a coarse prediction model, which is used to control the direction of data stream in this model. The output predicted by $D$ determines the fine prediction models the data stream will enter. That is, the results of model $D$ decide that our model will choose which fine prediction models in the second layer to accomplish the final prediction. Meanwhile, the high-layer sub-models closed to the output layers which contain $S_1$ to $S_n$ are called the fine prediction models. Those sub-models are used to do the exact predictions in parallel.

We use the latitude and longitude coordinates of base stations as the exact position coordinates in our mobile data. People can just appear in the base station they are communicating with near by due to the limit of communication range of base stations. Each base station is set to a label of our prediction model. Those labels are the final prediction labels \{$L_1$, $L_2$, ..., $L_n$\} for our model. We divide those labels into fine prediction models \{$S_1$, $S_2$, ..., $S_n$\} as the fine prediction labels for those sub-models according to the hierarchical spatial feature in our dataset (we will talk about it in detail in Section 5). When the data stream entry to one of fine models, we take the output in this fine model as the final prediction result. In this way, a fine prediction model only reach several final prediction labels. For example, the fine prediction model $S_1$ may be only used to predict the labels \{$L_1$, $L_2$\}.

Specifically, when the mobile data $M$ is coming, the data stream will arrive in the model $D$ firstly. This coarse prediction model $D$ is also a CNN, whose labels are named as ${L_{area}}$. In our task, the coarse prediction model is used to do the crude predictions, which means that we firstly locate a man in a crude range of area. As mentioned above, we gather some final labels \{$L_1$, $L_2$, ..., $L_n$\} into new classes, which are just the new labels in ${L_{area}}$. Then after processed by model $D$, the data stream will be directed by the labels ${L_{area}}$. Data will entry to the correspond fine prediction model in \{$S_1$, $S_2$, ..., $S_n$\}. In this process, we can run our algorithm in parallel. When new data is coming, the model can be trained at the same time if the label ${L_{area}}$ of new data is different from the label ${L_{area}'}$ of past data because those fine models are all independent. Therefore, there are the two levels of hierarchy in ''DeepSpace", the first is $D$, and the second are the models $S_1$ to $S_n$. This is illustrated in Algorithm \ref{alg:example1}.

\begin{algorithm}[htb]
\caption{DeepSpace Model} 
\label{alg:example1}  
\begin{algorithmic}
\Require the mobile data = \{$M$\} , data's crude area label = \{${L_{area}}$\}, data's exact position label = \{$L_1$, $L_2$, ..., $L_n$\} and  the current sub-models = \{$D$, $S_1$, $S_2$, ..., $S_n$\}
\Ensure the updated models = \{$D$, $S_1$, $S_2$, ..., $S_n$\}

train the $D$ using the input data $M$ to gain the coarse prediction label ${L_{area}}$

/* parallel computation part */

\For {all $l \in \{L_1, L_2, ..., L_n\}$ and $s \in \{S_1, S_2, ..., S_n\}$}
    \If {${L_i} \in {L_{area}}$}
        \State lead to data stream $M$ to the fine prediction model $S_i$ corresponding to the exact position label $L_i$, and train $S_i$ to update its parameters
    \EndIf
\EndFor

\Return \{$D$, $S_1$, $S_2$, ..., $S_n$\}
\end{algorithmic}
\end{algorithm}

\section{Dataset and Analysis}
The dataset used for our experiment is collected from the mobile cellular network in a city of southeastern China, which contains many different usage detail records (UDRs) of mobile communication. It is provided by the China Mobile Communications Corporation, which is only for the scientific research and forbidden for commercial use. This dataset is not open-accessed because the UDRs contains too much private individual information such as the locations of users, the contents that users browse and so on. In this section, we will introduce our dataset in detail, and analyze the characteristics of this dataset.

\subsection{Dataset}
\textbf{Details of dataset.} When predicting human trajectories via mobile big data, the call detail records (CDRs) from the mobile cellular network have been widely used by many researchers \cite{CDR1,CDR2,CDR3}. Specially, with the limitation that CDR contains little spatial information, we consider using the data usage detail records (UDRs) to track human movement. The UDRs contains more information about the human movement pattern due to rich records toward human activities. As a result, this dataset is more suitable to be used to study the spatial collaboration in the mobile cellular network.

The dataset on UDRs in our experiments is collected in Jinhua City, located in the middle of Zhejiang Province, China. It contains of data access records made by the crowd within this area, covering over two thousands base stations, and is collected from November 21st, 2014 to December 13rd, 2014. The UDR provides a detailed description of human movement. In our dataset, a UDR contains 40 fields to describe the human mobile communication activities in the mobile cellular network. However, some fields may reflect the similar information (such as the phone numbers and terminal numbers of users), or may be indifferent for predicting human trajectories (such as the server ip users request when surfing the Internet through mobile devices). We show nine fields in Table \ref{table_data}. In our experiment, we will focus on the fields in this Table \ref{table_data} to train our ``DeepSpace" model because we believe those fields can represent the characteristics of human movement well.

\begin{table}[!t]
\newcommand{\tabincell}[2]{\begin{tabular}{@{}#1@{}}#2\end{tabular}}
\renewcommand{\arraystretch}{1.3}
\caption{Fields of our dataset}
\label{table_data}
\centering
\begin{tabular}{|c|c|}
\hline
Fields & Description\\
\hline
phonenum & \tabincell{c}{an encrypted telephone number\\ indicating a mobile user}\\
\hline
stime & the beginning time of a data usage record\\
\hline
etime & the end time of a data usage record\\
\hline
host & \tabincell{c}{the domain name that a user requests\\ in a data usage record}\\
\hline
appid & the application number in a data usage record\\
\hline
url & \tabincell{c}{the Internet address that a user requests\\ in a data usage record}\\
\hline
lacid & the location area code (LAC) of a data usage record\\
\hline
longitude & \tabincell{c}{ the longitude of the base station \\ in a data usage record}\\
\hline
latitude & the latitude of the base station in a data usage record\\
\hline
\end{tabular}
\end{table}

\begin{figure}[tbp]
 \subfigure[\footnotesize Geographic Map]{
 \label{fig:location:a}
\includegraphics[height=4cm, width=4cm]{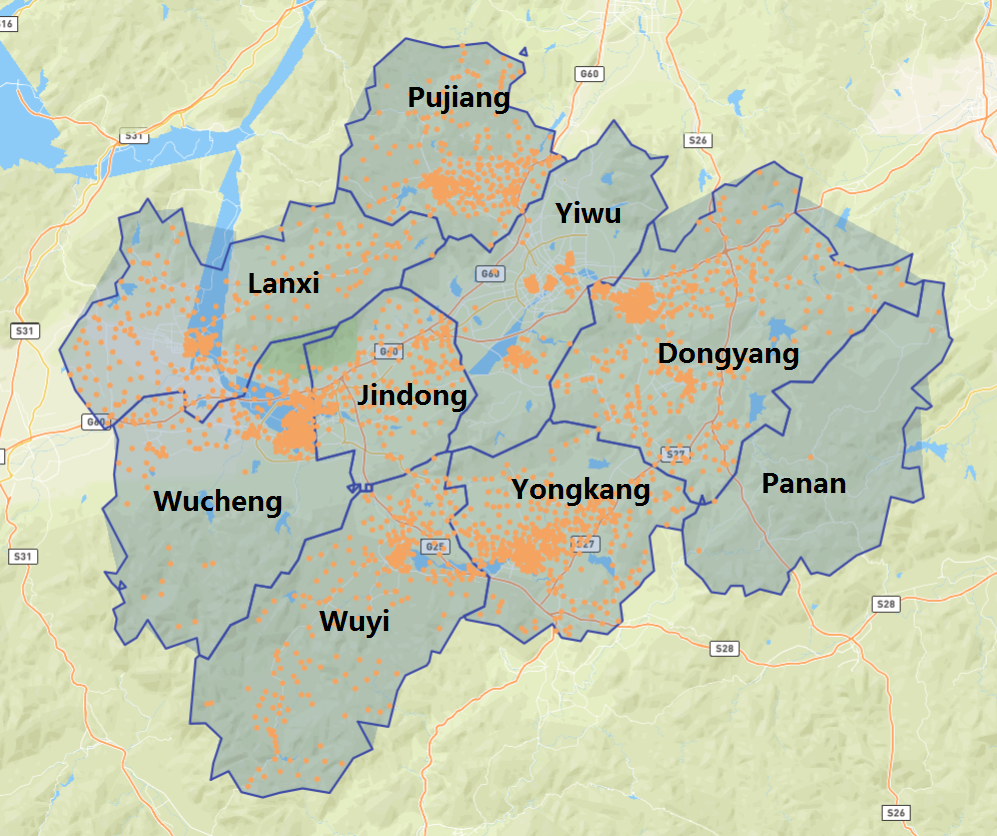}}
\hspace{0.001in}
 \subfigure[\footnotesize Voronoi Map]{
  \label{fig:location:b}
  \includegraphics[height=4cm, width=4.5cm]{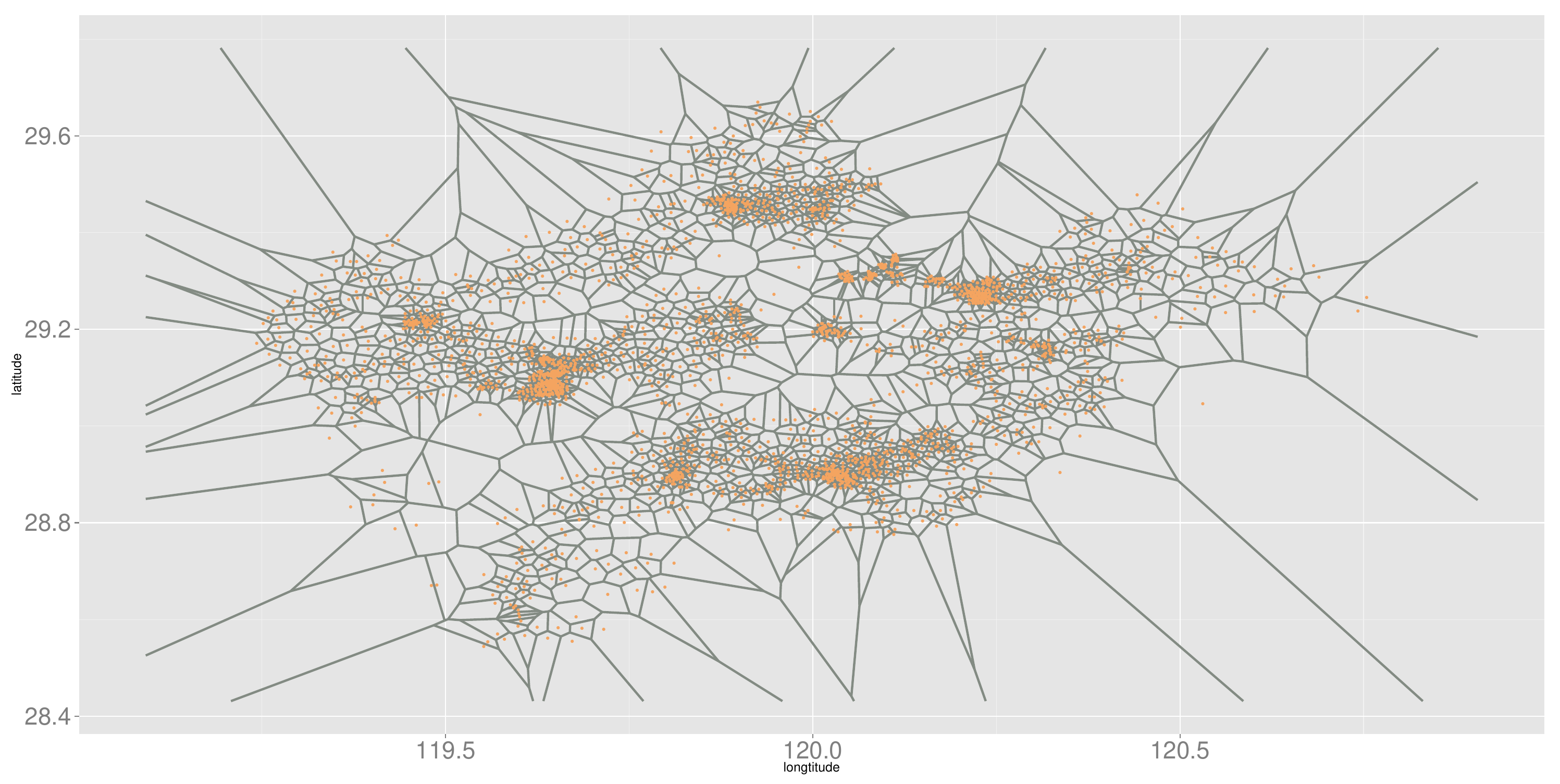}}
 \caption{ (a) and (b) : the distribution of base stations on geographic map and voronoi map. In (b), the x-coordinate represents longitude and y-coordinate represents latitude.
}
 \label{fig:subfig}
\end{figure}

When a user accesses the mobile Internet using the mobile device via a base station, this user's UDRs (e.g., duration, URL etc.) will be recorded by the base station in this user's present position. We use the location of base stations a user communicates with to represent his/her the fine location. In Fig. \ref{fig:location:a}, we show the distribution of base stations in Jinhua City, and the territory of counties and cites belongs to Jinhua. In this geographic map, there are less base stations in the remote country area because of lower population density and slower population mobility when compared with the town. Therefore, the coverage of base stations in the remote country can not be the same as those in the town. To consider this factor, we use the voronoi map (Fig. \ref{fig:location:b}) to redraw this distribution of base stations in Jinhua. In Fig. \ref{fig:location:b}, the zones of base station signal coverage are divided into independent area element (the x-coordinate represents longitude and y-coordinate represents latitude). It is straightforward that the coverage of base stations in the remote country area is bigger.

\textbf{Analysis of dataset.} We focus on the spatial features in this dataset to analyze the human movement. We notice the ``lacid" field, ``longitude" and ``latitude" fields consist of hierarchical spatial association. The idea of ``DeepSpace" model is exactly from this hierarchical association. The ``lacid" field refers to the location area code (LAC) in mobile communication system. LAC is designed for paging service, covering a geographic area. It is divided by administrative areas (a county or district). When we use the location of base stations to represent positions of users, LAC indicates the coarse area users belong to but longitude and latitude refers to the precise location. This spatial feature corresponds to our ``DeepSpace" model perfectly. Specifically, ``lacid" corresponds to labels ${L_{area}}$ of the coarse prediction model $D$, while ``longitude" and ``latitude" corresponds to labels \{$L_1$, $L_2$, ..., $L_n$\} of the fine prediction model \{$S_1$, $S_2$, ..., $S_n$\} in Section 4.

\begin{figure}[tbp]
 \subfigure[\footnotesize LAC]{
 \label{fig:subfig:a}
\includegraphics[height=2.5cm, width=8.5cm]{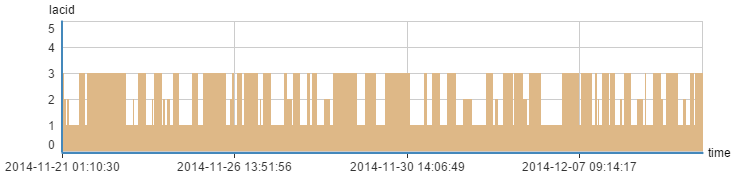}}
 \subfigure[\footnotesize Longitude and latitude]{
  \label{fig:subfig:b}
  \includegraphics[height=2.5cm, width=8.5cm]{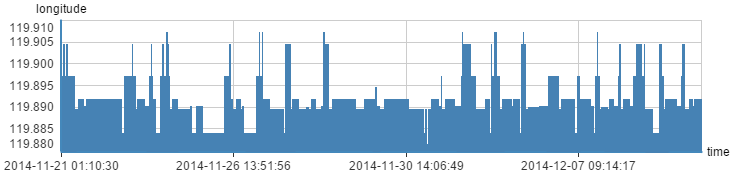}}
 \caption{ (a) and (b) : Changes of positions in two spatial scale: LAC, longitude and latitude.}
 \label{fig:change}
\end{figure}

Meanwhile, we find the changes of ``lacid" are more periodic than changes of ``longitude" and ``latitude" when analyzing our dataset. In Fig. \ref{fig:change}, we can see the changes curve of ``lacid" per day is similar, while the curve of ``longitude" and ``latitude" is less periodic. This means that we can make a more accurate prediction via learning the daily feature of position changes in the scale of ``LAC".

\subsection{Data Cleaning}
Before we use this dataset to do experiments, this data need to be cleaned to remove the significant noise. There exists some ``dirty data" in our dataset because of mistakes in the data recording or collecting systems. Some data usage records are redundant or incomplete (some important fields like ``stime" and ``etime" is empty). We remove those ``dirty data" from our dataset.

\begin{figure}[!htb]
\centering
\includegraphics[height=3cm, width=5.4cm]{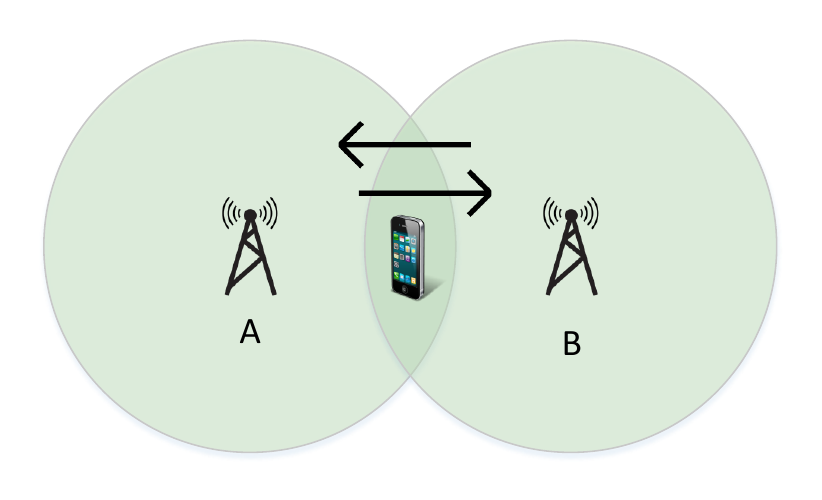}
\caption{Base stations switching phenomena.}
\label{fig:switch}
\vspace{-.8em}
\end{figure}

Furthermore, what we are more concerned about in data cleaning is an interesting phenomena (base stations switching phenomena). Because we use the locations of base stations to represent the positions of users, the relationship between locations of base station and positions of users should be reliable. However, there exist some strange records like those in Table \ref{table_clear1}. We can see the ``etime" in the first record and the ``stime" in the second record is the same, while the longitude and latitude are different in those two records. People can not show up in two places at the same time. This strange situation is caused by the base station switching phenomena. As be shown in Fig. \ref{fig:switch}, there always exist some overlapping area between two base station ($A$ and $B$ in this figure) coverage. When a user accesses the mobile internet in the overlapping area, this user may switch to the base station $B$ suddenly not on his/her own while this user always communicate with base station $A$ before. This phenomena happens due to changes of communication environment (such as changes of signal strength and base station's capacity). For those two records, we use the longitude and latitude position in the first record to replace the second one.

Another similar situation also caused by base stations switching phenomena is shown in Table \ref{table_clear2}. This user travels 68 kilometers inferred from the two records, while the time interval between the two records is less than 10 minutes. It is not possible for a person to move such a long distance in this short time. The reason is that this user travel in the overlapping area in this time interval. To address this problem, we first reorder the data records in time sequence. Let $T_i$ and $T_j$ be the ``etime" in the first record and ``stime" in the second record among two contiguous records. Let $lon_i$, $lat_i$ and $lon_i$, $lat_j$ denote the ``longitude" and ``latitude" in two contiguous records ($lon_i$, $lat_i$, $lon_i$ and $lat_j$ are expressed in radians). We can calculate the user's traveling speed $V_{ij}$ among the two contiguous records, where $R=6371$ kilometers indicates the earth's radius:
\begin{equation}
{V_{ij}} = \frac{{{D_{ij}}}}{{{T_j} - {T_i}}},
\end{equation}
where
\begin{equation}
\begin{split}
{D_{ij}} = R\arccos (\cos (la{t_i})\cos (la{t_j})\cos ((lo{n_j} - lo{n_i}))\\
 + \sin (la{t_i})\sin (la{t_j})).
\end{split}
\end{equation}
$D_{ij}$, the traveling distance among the two contiguous records, can be calculated by the longitude and latitude of base stations. If the traveling speed of users exceeds the maximum limitation ($V_ij$), we will treat those records as the base stations switching phenomena and use the longitude and latitude position in the first record to replace the second one.
\begin{table}[!t]
\newcommand{\tabincell}[2]{\begin{tabular}{@{}#1@{}}#2\end{tabular}}
\renewcommand{\arraystretch}{1.3}
\caption{Strange record a}
\label{table_clear1}
\centering
\begin{tabular}{|c|c|c|c|c|}
\hline
stime & etime & phonenum & longitude & latitude\\
\hline
\tabincell{c}{2014-11-26\\ 10:54:31} & \tabincell{c}{2014-11-26\\ 10:54:32} & 73913461166 & 119.90042 & 28.88195\\
\hline
\tabincell{c}{2014-11-26\\ 10:54:32} & \tabincell{c}{2014-11-26\\ 10:54:51} & 73913461166 & 119.89141 & 28.87161\\
\hline
\end{tabular}
\end{table}

\begin{table}[!t]
\newcommand{\tabincell}[2]{\begin{tabular}{@{}#1@{}}#2\end{tabular}}
\renewcommand{\arraystretch}{1.3}
\caption{Strange record b}
\label{table_clear2}
\centering
\begin{tabular}{|c|c|c|c|c|}
\hline
stime & etime & phonenum & longitude & latitude\\
\hline
\tabincell{c}{2014-11-24\\ 09:49:41} & \tabincell{c}{2014-11-24\\ 09:49:49} & 74424106409 & 120.07602 & 29.49888\\
\hline
\tabincell{c}{2014-11-24\\ 09:58:08} & \tabincell{c}{2014-11-24\\ 09:58:13} & 74424106409 & 120.04997 & 28.88697\\
\hline
\end{tabular}
\end{table}

\section{Experiment Evaluation}
We test our algorithm in our mobile dataset, and compare ``DeepSpace" model with the vanilla CNN. All the deep learning based algorithms are performed on a server equipped with a 2.50 GHz Intel(R) Xeon(R) E5-2680 CPU, a NVIDIA(R) Tesla(TM) K80 GPU and 64G RAM. We run our experiment via the open-source framework Caffe \cite{caffe}.

In this paper, Our experiments are based on the data that is collected in the twenty three days of November and December in 2014. The data of the first nineteen days is treated as our training set, while the remaining data collected in the last four days is used for testing.

\subsection{Determination of the Structure of ``DeepSpace"}
As mentioned above, ``DeepSpace" consists of the coarse prediction model and fine prediction models (all those sub-models are CNNs). To develop our model, we need to determine the size of the input layer, the number of hidden layers, the layers structure and the number of hidden units in both the coarse predicting model and fine predicting models. For the input layer, considering the temporal correlation of human trajectories, we use the historical position at the previous time intervals as the input data to prediction the position at present time interval. In fact, the length of previous time intervals affects the accuracy of our prediction. In this article, we choose 4 different length of time intervals to analyse the example results from \{50, 100, 150, 200\}.

\begin{figure}[!htb]
\centering
\includegraphics[height=5.5cm, width=6.5cm]{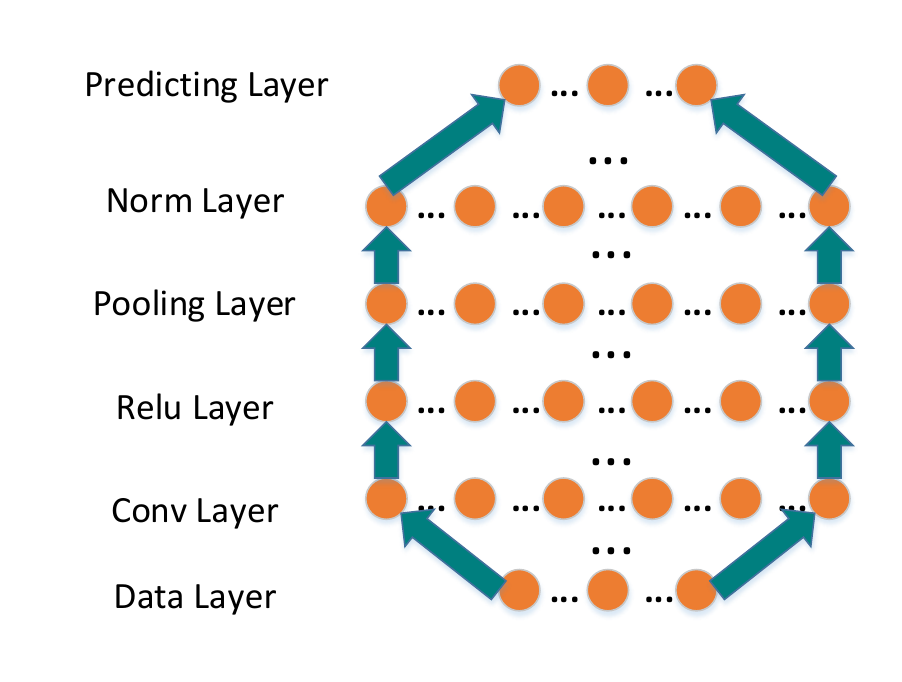}
\caption{The architecture of CNN. }
\label{fig:cnn}
\vspace{-.8em}
\end{figure}

After numerous experiments, we obtain the best CNN architecture, which is shown in Fig. \ref{fig:cnn}. This architecture consists of four hidden layers, include convolutional layers (Conv layer), Parametric Rectified Linear Unit layers (PReLU layer) \cite{Ch}, max-pooling layers (Pooling layer) and normalization layers (Norm layer). According to our experiments, we find this architecture performs well as both the coarse prediction model and fine prediction models. This phenomenon hints to the spatial correlations behind the data of our coarse prediction model and fine prediction models.

The historical position sequence data is processed by a pair of convolutional layer and max-pooling layer. Then the first convolutional layer is followed by a PReLU layer. The convolutional layer has 25 hidden units. Next, there is a PReLU layer in order to improve model fitting with nearly zero extra computational cost and reduce the risk of overfitting. Then, there is the max-pooling layer in order to compromise the features. Furthermore, the last layer of the network is the predicting layer. This layer is designed to learn the parameter $w$ by minimizing the loss function ($J(w)$) using a softmax classifier (${x^{(i)}}$ and ${y^{(i)}}$ are the feature vector and label for the $i$-th instance respectively):
\begin{equation}
J(w) =  - \frac{1}{m}[\sum\limits_{i = 1}^m {\sum\limits_{j = 0}^1 {1\{ {y^{(i)}} = j\} \log p({y^{(i)}} = j|{x^{(i)}};w )} } ],
\end{equation}
where
\begin{equation}
p({y^{(i)}} = j|{x^{(i)}};w) = \frac{{{e^{w_j^T{x^{(i)}}}}}}{{\sum\nolimits_{l = 1}^k {{e^{w_j^T{x^{(i)}}}}} }}.
\end{equation}

Extrapolating from our experiment results, we find all sub-models in ``DeepSpace" can not have too many layers. We indicate the reason is that if we use deeper sub-models, ``DeepSpace" will be overfitting because of huge parameters of complex sub-models when ''DeepSpace" is complex hierarchial structure. Besides, in our experiments, we find that setting the bias of the convolution layer to $0.9$ can do results in better performance instead of initializing it to $0$ in most experiments. This simple adjustment can improve about nearly $1\%$ accuracy rate compared to setting this to $0$.

\subsection{Result}
There are two different sub-models to train in our model. These two different sub-models are trained independently in our experiments. We both enter the historical position sequence data into these models. However, the input of the coarse prediction model is the whole training data (the whole nineteen days' data), but the input of the fine prediction models is the selected data from the whole data. We use the historical position at the previous time intervals (training data) to predict the position at present time interval (the label corresponding to this example of training data). In this paper, we use 50-time-intervals sequences, 100-time-intervals sequences, 150-time-intervals sequences, and 200-time-intervals sequences respectively to train our proposed model. The 50-time-intervals sequences means the 50 contiguous records in order of time occurred in our dataset. The rest is similar. When the position at present time interval (the label) is in the range of labels of fine models in \{$S_1$, $S_2$, ..., $S_n$\}, this example of training data will be divided into the training set of this fine model. By repeating this algorithm, we divide the whole training data set into many part to fit those fine models.

Although we use the whole nineteen days' data to train coarse model, we make a change in the encoding mode of the historical position sequence. For the training data in the fine predicting models, we use the numbers in \{0, 1, ..., $n - 1$\} ($n$ is the number of different fine positions) to encode the historical positions sequence. On the contrary, we use the numbers in \{0, 1, ..., $n' - 1$\} ($n'$ is the number of different ``lacid"). Users' movement is more regular on the coarse scale of ``lacid" than the fine position (accurate to longitude and latitude). So the way we use the coarse scale map to recode the training data help to filtrate the noise bring by the too fine positions information. As a result, this adjustment brings an incredible improvement for the coarse predicting accuracy about $30\%$ on the average.

\begin{table*}[!t]
\caption{Accuracies of ''DeepSpace" and the vanilla CNN}
\label{accuracy}
\centering
\begin{tabular}{|c|c|c|c|c|}
 \hline
 \multirow{2}{*}{Model} &
 \multicolumn{3}{c|}{DeepSpace} &
 \multirow{2}{*}{vanilla CNN} \\
 \cline{2-4}
   & coarse predicting model & fine predicting models & the whole model & \\
 \hline
 50-time-intervals & 86.18\% & 86.92\% & 71.01\% & 62.44\% \\
 \hline
 100-time-intervals & 85.83\% & 86.9\% & 70.02\% & 59.78\% \\
 \hline
 150-time-intervals & 87.41\% & 87.53\% & 71.29\% & 58.95\% \\
 \hline
 200-time-intervals & 85.17\% & 87.78\% & 70.21\% & 56.24\% \\
 \hline
 \end{tabular}
 \end{table*}

\begin{figure*}[tbp]
 \subfigure[A user who moves frequently.]{
 \label{fig:subfig:u1}
\includegraphics[height=5cm, width=9cm]{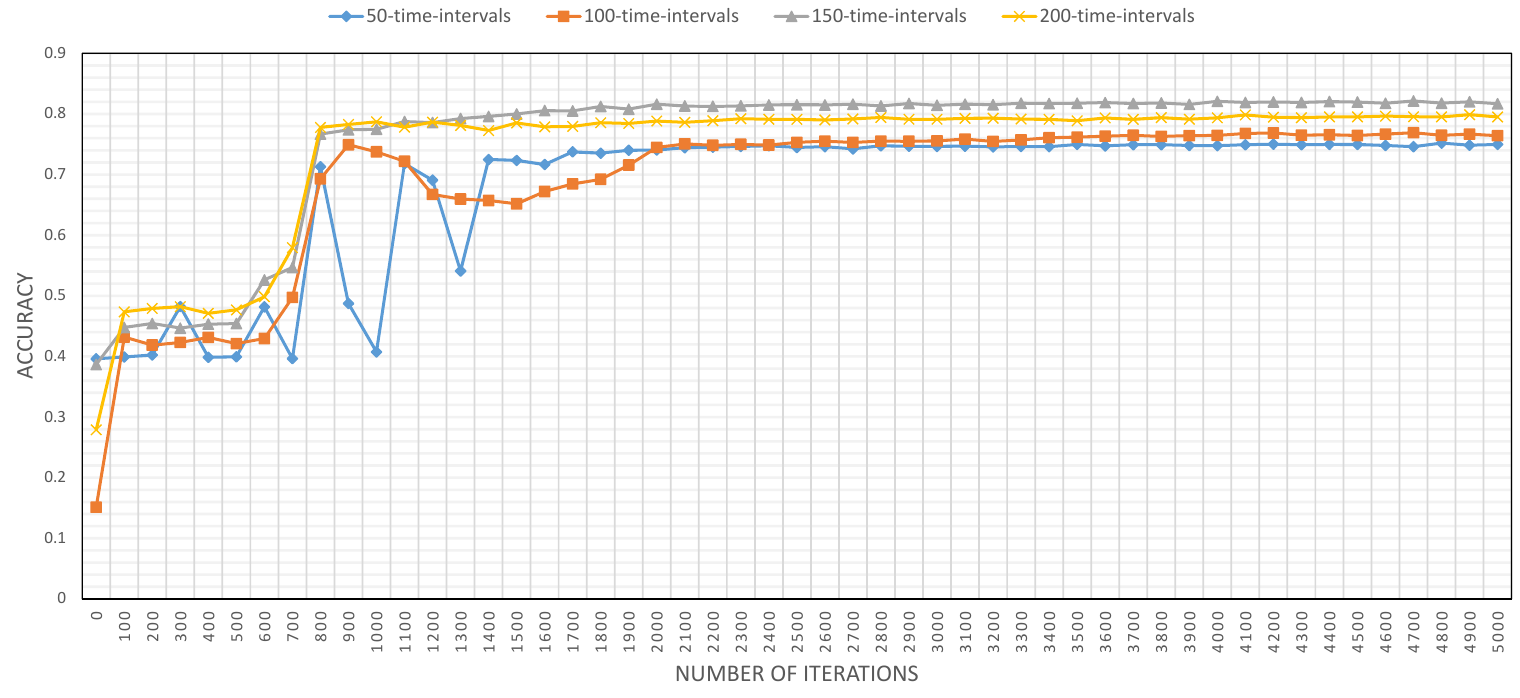}}
\hspace{0.001in}
 \subfigure[A user who moves less frequently]{
  \label{fig:subfig:u2}
  \includegraphics[height=5cm, width=9cm]{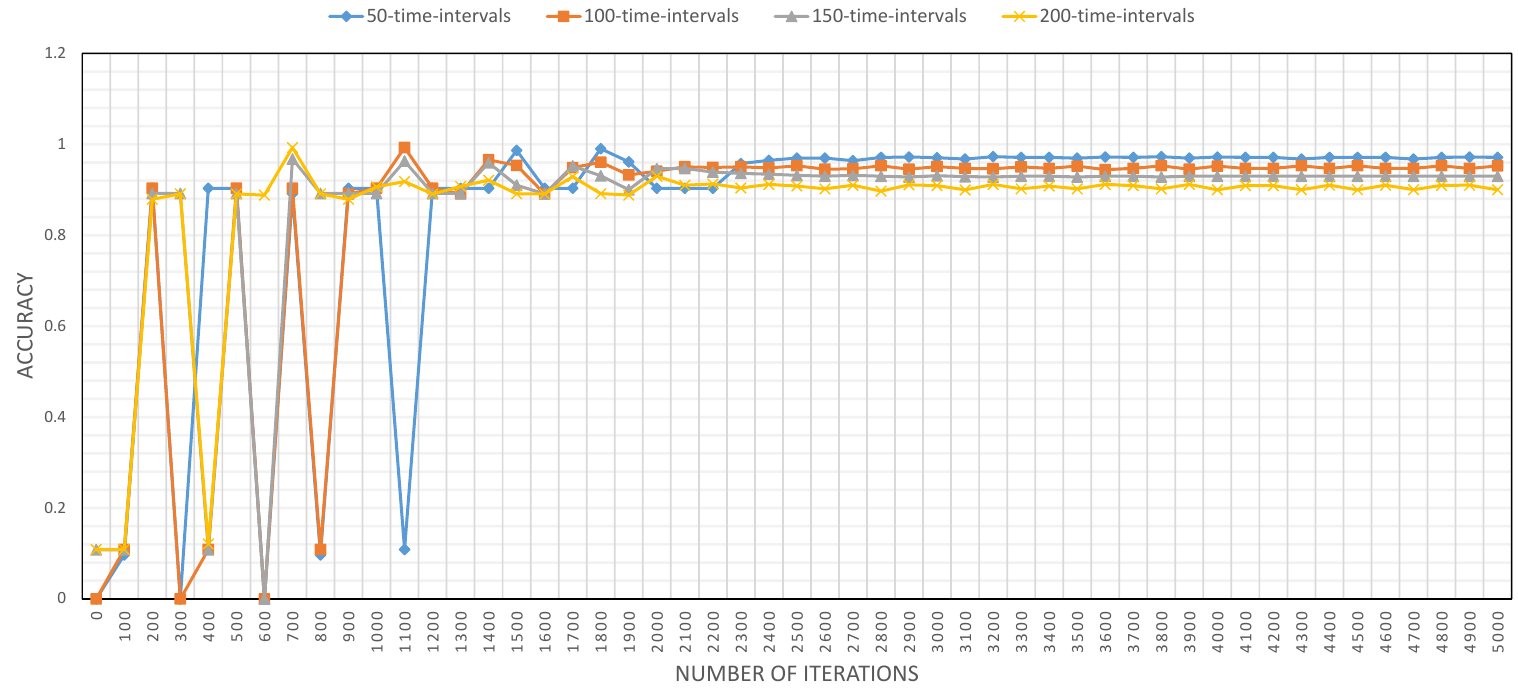}}
 \caption{ (a) and (b): The training curve of two users via ``DeepSpace".
}
 \label{fig:subfig}
\end{figure*}

Fig. \ref{fig:subfig} presents the processes of training of the coarse prediction models for the human mobility of two typical users with more and less mobility. When we train ``DeepSpace" model, the coarse prediction model is much more important because the number of the base station that a user will arrive in the range of a ``lacid" (the coarse positions) is limited. Specifically, the key influence to the accuracy of our model is the accuracy of the coarse prediction model. Then, we do experiments on users who have different behavioral idioms. In Fig. \ref{fig:subfig}, every curve represents the convergence of the accuracy of training. Fig. \ref{fig:subfig:u1} shows the accuracy variation of a user who moves more frequently. Instead, the change of accuracy of a user who moves less frequently is shown on the fig. \ref{fig:subfig:u2}. We compare the processes of training for 50, 100, 150, 200 time intervals aforementioned in both two graphs. When we analyze the convergence curves of different time intervals, the curvilinear trend is similar for each user. However, the numbers of iterations are obviously different when the accuracy curve is converged. As shown in Fig. \ref{fig:subfig:u1} and Fig. \ref{fig:subfig:u2}, the model converge faster with more time intervals. Before the curve becomes converged, it is a wiggly line with growth movement especially in Fig. \ref{fig:subfig:u2}. From Fig. \ref{fig:subfig:u1} and Fig. \ref{fig:subfig:u2}, we can clearly see that more previous time intervals for prediction will not necessary lead to better accuracy. In Fig. \ref{fig:subfig:u1}, 150-time-intervals gain the best accuracy and it's 50-time-intervals in Fig. \ref{fig:subfig:u2}. Furthermore, our model gains the best accuracy of 82.09\% for the user who moves frequently in Fig. \ref{fig:subfig:u1}, while the best accuracy is 97.2\% for the user who moves less frequently in Fig. \ref{fig:subfig:u2}. This result is scrutable if we consider that the former moves more frequently. The more changes lead to more difficulties in predicting human trajectories, which will generates more noise than the latter. The former user has been to twelve base stations while the latter has only been to four base stations. In spite of the frequent movement, the coarse prediction model still gains the accuracy of 82.09\%.

\begin{figure}[!htb]
\centering
\includegraphics[height=5cm, width=9cm]{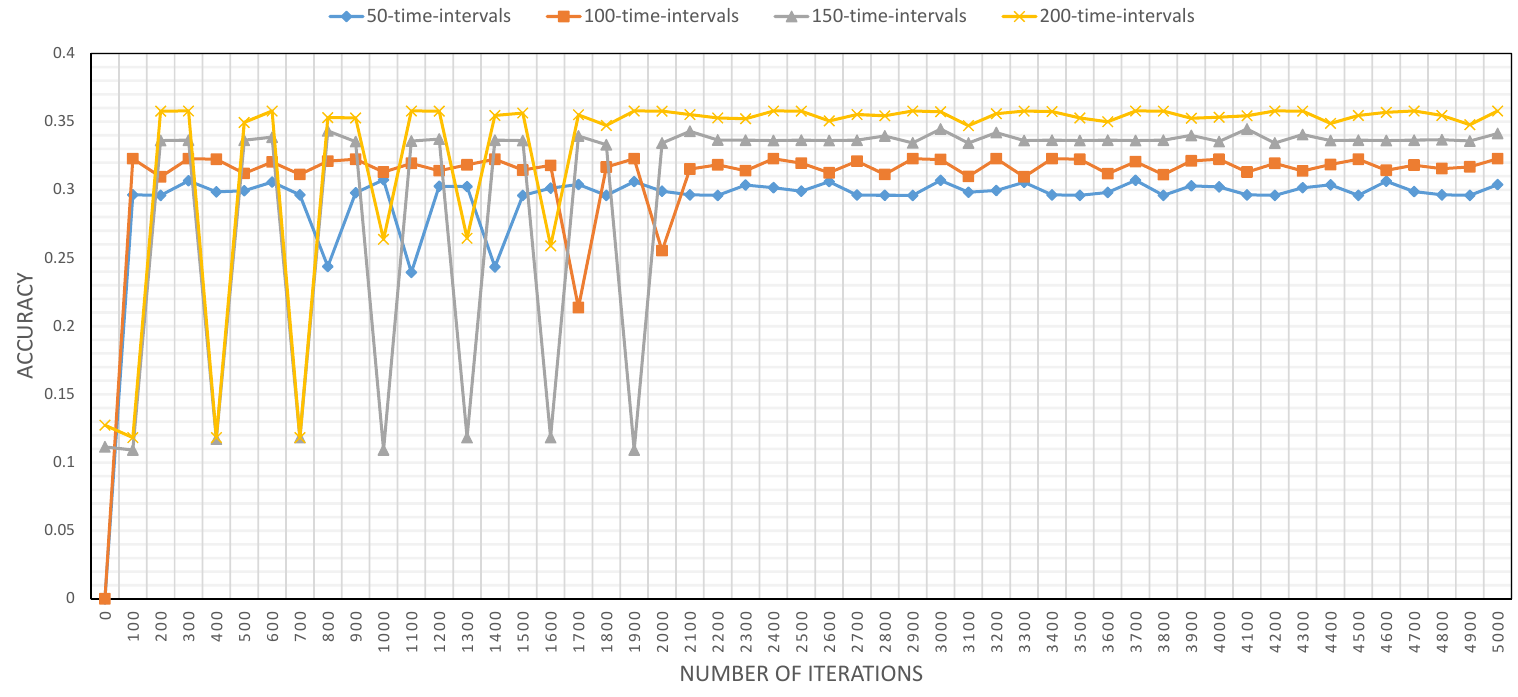}
\caption{The training curve of the vanilla CNN. }
\label{fig:vanilla}
\vspace{-.8em}
\end{figure}

\begin{figure*}[tbp]
 \subfigure[The user who moves less frequently.]{
 \label{fig:path:u1}
\includegraphics[height=4.9cm, width=7.4cm]{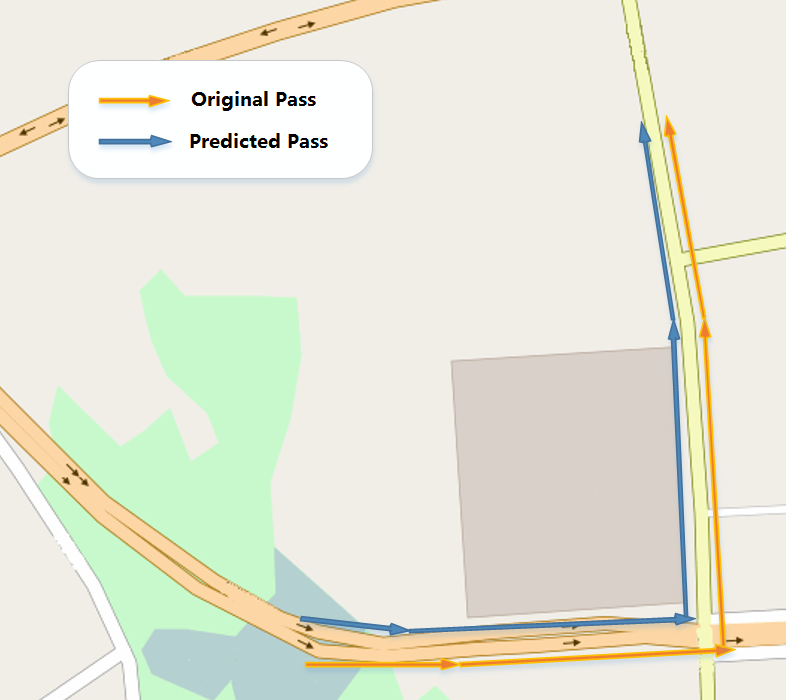}}
\hspace{0.6in}
 \subfigure[The user who moves more frequently.]{
  \label{fig:path:u2}
  \includegraphics[height=5cm, width=7.5cm]{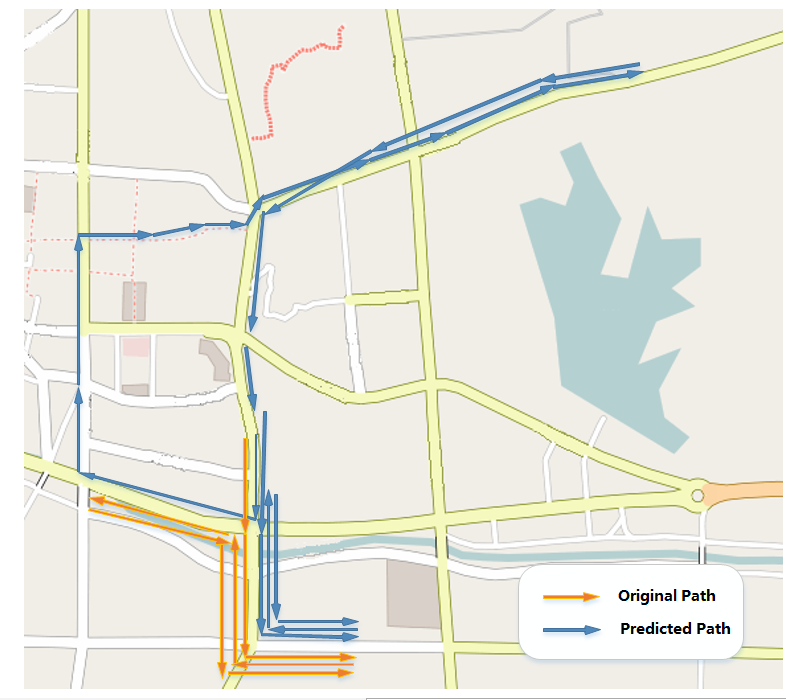}}
 \caption{ (a) and (b): The original and predicted paths of two users via ``DeepSpace".}
 \label{fig:path}
\end{figure*}

We compare the performance of the vanilla CNN model and ``DeepSpace" model. Fig. \ref{fig:vanilla} shows the convergence of training of the vanilla CNN model. Similarly with training ``DeepSpace" model, we also use the historical position at the previous time intervals to prediction the position at present time interval. Also, we consider different length of time (50 time intervals, 100 time intervals, 150 time intervals, and 200 time intervals) to do the prediction. From Fig. \ref{fig:vanilla}, we can see that we get higher accuracy using more previous time intervals to do the prediction. However, it is not always the same. Fig. \ref{fig:vanilla} is the result of the user who move frequently (the user in Fig. \ref{fig:subfig:u1}). But for some other users, situation may be diametrically opposed. We can tell the best accuracy in fig. \ref{fig:vanilla} is 35.77\%, which far less than both Fig. \ref{fig:subfig:u1}) and Fig. \ref{fig:subfig:u2}). More details of the comparison of ``DeepSpace" and the vanilla CNN is shown in Table \ref{accuracy}. This table shows the average accuracies (including the coarse predicting model, fine predicting models and the whole model of ``DeepSpace") in our mobile dataset. Our network model makes a show of strength compared to the vanilla CNN. In spite of any number of time intervals, ``DeepSpace" performs much better than the vanilla CNN. ``DeepSpace" gains over 85\% of accuracy both for the coarse predicting model and fine predicting models in our dataset. After the hierarchical structure, our model gains approximately 70\% of accuracy, which is almost a 10 percent drop compared to the accuracy of the vanilla CNN. Furthermore, our model achieves 71.29\% in the 150 time intervals, which is the best performance. In fact, ``DeepSpace" performs stably in different time intervals. However, the accuracies of the vanilla CNN in different time intervals falls off quickly as the number of time intervals increases. The reason is that more intervals lead to more noise. Using less previous time intervals to do predicting means a simpler patten to learning for the models. This phenomena indicates human mobility patterns more depend on the recent history. ``DeepSpace" has high ability of tolerating noise with benefit of the hierarchical structure. Actually, the vanilla CNN performs much worse than ``DeepSpace" for those user who moves frequently. For instance, the vanilla CNN only achieves 35.77\% accuracy for the user in Fig. \ref{fig:subfig:u1}) who moves frequently in the best of experiment results. However, ``DeepSpace" gain 57.83\% accuracy for this user.

We show our predicting moving path and the real moving path in the real geographic map in Fig. \ref{fig:path}. We establish the last four days (those four days' data is our test dataset of experiments as mentioned before) of two users. In Fig. \ref{fig:path}, we distinguish the the original path and predicted path with different color lines (one is denoted by orange lines while another is denoted by blue lines). We can find the original path coincides with the predicted path basically in Fig. \ref{fig:path:u1}. Considering that the accuracy of our model for this user is 97.2\%, this result meets our expectation. However, in Fig. \ref{fig:path:u2}, the original path only coincides with the predicted path in the lower portion of the picture well. In the higher portion of the picture, the orange lines even does not appear while blue lines exist. We find this user only walk through the blue lines in the higher portion of the picture only once time in our 4-days dataset when we observe the data of this user. In Fig. \ref{fig:path:u2}, ``DeepSpace" can capture the pattern for those dense lines in the graph well. If a user changes a normal pattern and goes to a place suddenly, ``DeepSpace" may ignore this event.

\section{Conclusion}
We propose an online deep learning framework for predicting human moving path. Considering that the CNN can deal with the time series data well, we transform the vanilla CNN to a hierarchial structure, including the coarse predicting model and fine predicting models. Thanks to this hierarchial structure, our algorithm can deal with data online and in parallel. ``DeepSpace" is able to extract the spatiotemporal features of our network-level mobile data. We evaluate the performance of the proposed method on the dataset collected from the mobile cellular network in a city of southeastern China, which contains many different usage detail records (UDRs) of mobile communication, and compare it with the vanilla CNN. In addition, we analyze the influence of different lengths of time intervals used for prediction. The results show that ``DeepSpace" is superior to the vanilla CNN.

Although ``DeepSpace" achieves a great performance in our experiment, there are still many interesting questions to think. In this paper, we capture the spatiotemporal features of our dataset via the CNN models. In our experiment, we try to use 50, 100, 150, and 200 previous time intervals to predict the present positions of users. For the vanilla CNN model, the accuracies of the vanilla CNN in different time intervals falls off quickly as the number of time intervals increases. As such, the CNN is hard to deal with the long time series data. For the future work, it would be interesting to combine the CNN model with other neural networks (like the Recurrent Neural Network \cite{rnn} or the Long-Short-Term-Memory architecture \cite{lstm}). For RNN, neurons share the same weights at different time steps, allowing us to apply the network to input sequences of different lengths. LSTM solve the problem of vanishing gradients in long-term dependencies \cite{rnnques1} \cite{rnnques2}. Those networks can be helpful to deal with the long time series data. This combination may lead to a more complex structure and more parameters to train, but we believe this models may make further performance improvement.


%

%

\ifCLASSOPTIONcompsoc
  \section*{Acknowledgments}
\else
  \section*{Acknowledgment}
\fi

This research is supported by NSFC No. 61401169.

\ifCLASSOPTIONcaptionsoff
  \newpage
\fi



%

%

\begin{IEEEbiography}[{\includegraphics[width=1in,height=1.25in,clip,keepaspectratio]{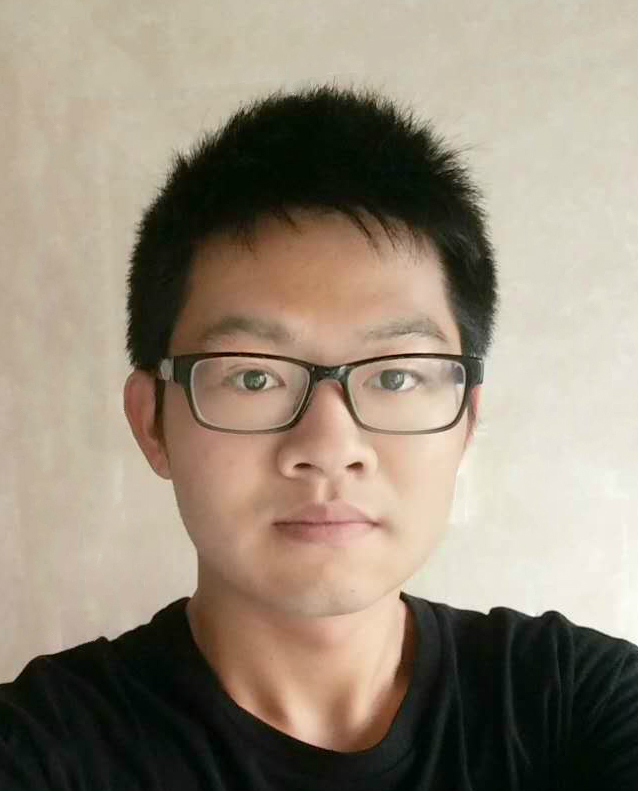}}]{Xi Ouyang}
received his B.S degree at the School of Electronic Information and Communications, Huazhong University of Science and Technology, Wuhan, P.R. China. He is currently working toward the M.S. degree still at the School of Electronic Information and Communications, Huazhong University of Science and Technology. His current research interests include: machine learning , deep learning and its applications.
\end{IEEEbiography}

\begin{IEEEbiography}[{\includegraphics[width=1in,height=1.25in,clip,keepaspectratio]{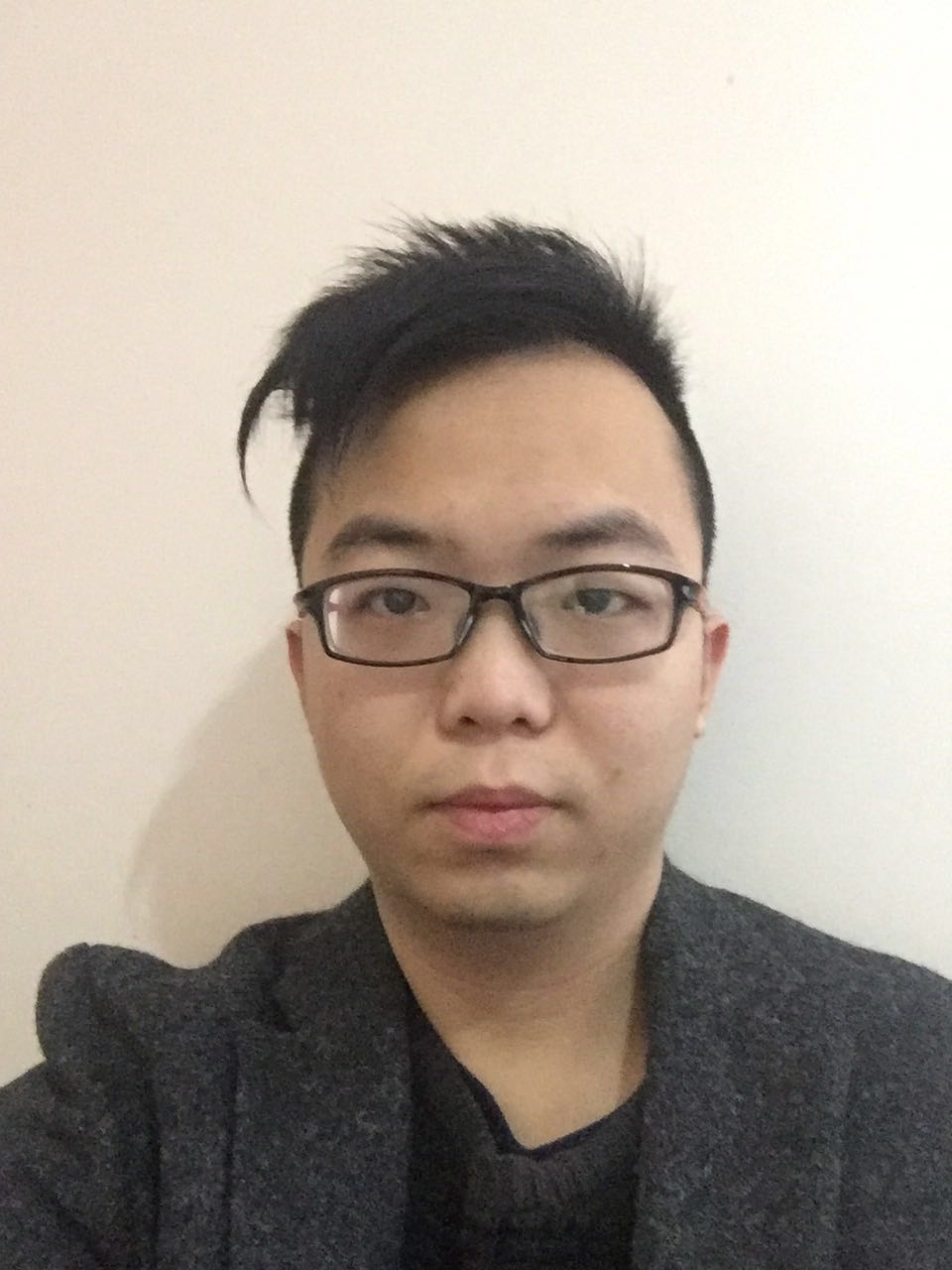}}]{Chaoyun Zhang}
received his B.S degree at the School of Electronic Information and Communications, Huazhong University of Science and Technology, Wuhan, P.R. China. He is currently working toward the Master degree at the School of Informatics, the University of Edinburgh, Edinburgh, UK.
His current research interests include: deep learning, mobile big data and pattern recognition.
\end{IEEEbiography}

\begin{IEEEbiography}[{\includegraphics[width=1in,height=1.25in,clip,keepaspectratio]{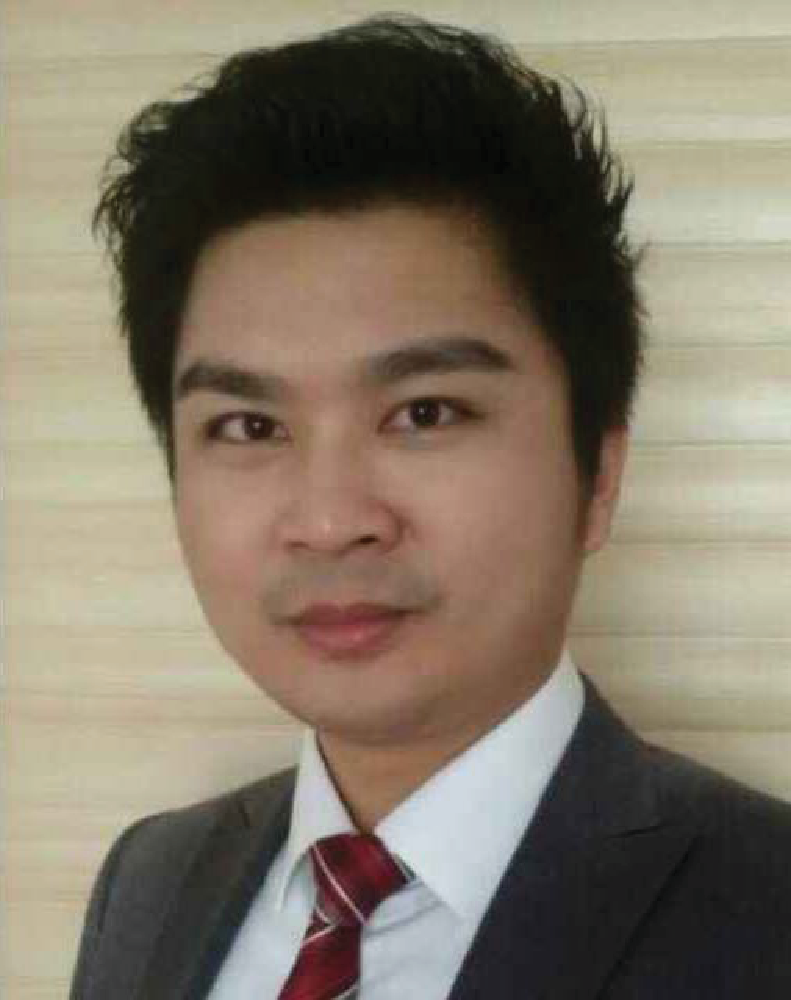}}]{Pan Zhou}
is currently an associate professor with School of Electronic Information and Communications, Huazhong University of Science and Technology,
Wuhan, P.R. China. He received his Ph.D. in the School of Electrical and Computer Engineering at the Georgia Institute of Technology (Georgia Tech) in 2011, Atlanta, USA. He received his B.S. degree in the Advanced Class of HUST, and a M.S. degree in the Department of Electronics and Information Engineering from HUST,Wuhan, China, in 2006 and 2008, respectively. He held honorary degree in his bachelor and merit research award of HUST in his master study. He was a senior technical member at Oracle Inc, America during 2011 to 2013, Boston, MA, USA, and worked on hadoop and distributed storage system for big data analytics at Oralce cloud Platform. His current research interest includes: communication and information networks, security and privacy, machine learning and big data.
\end{IEEEbiography}

\begin{IEEEbiography}[{\includegraphics[width=1in,height=1.25in,clip,keepaspectratio]{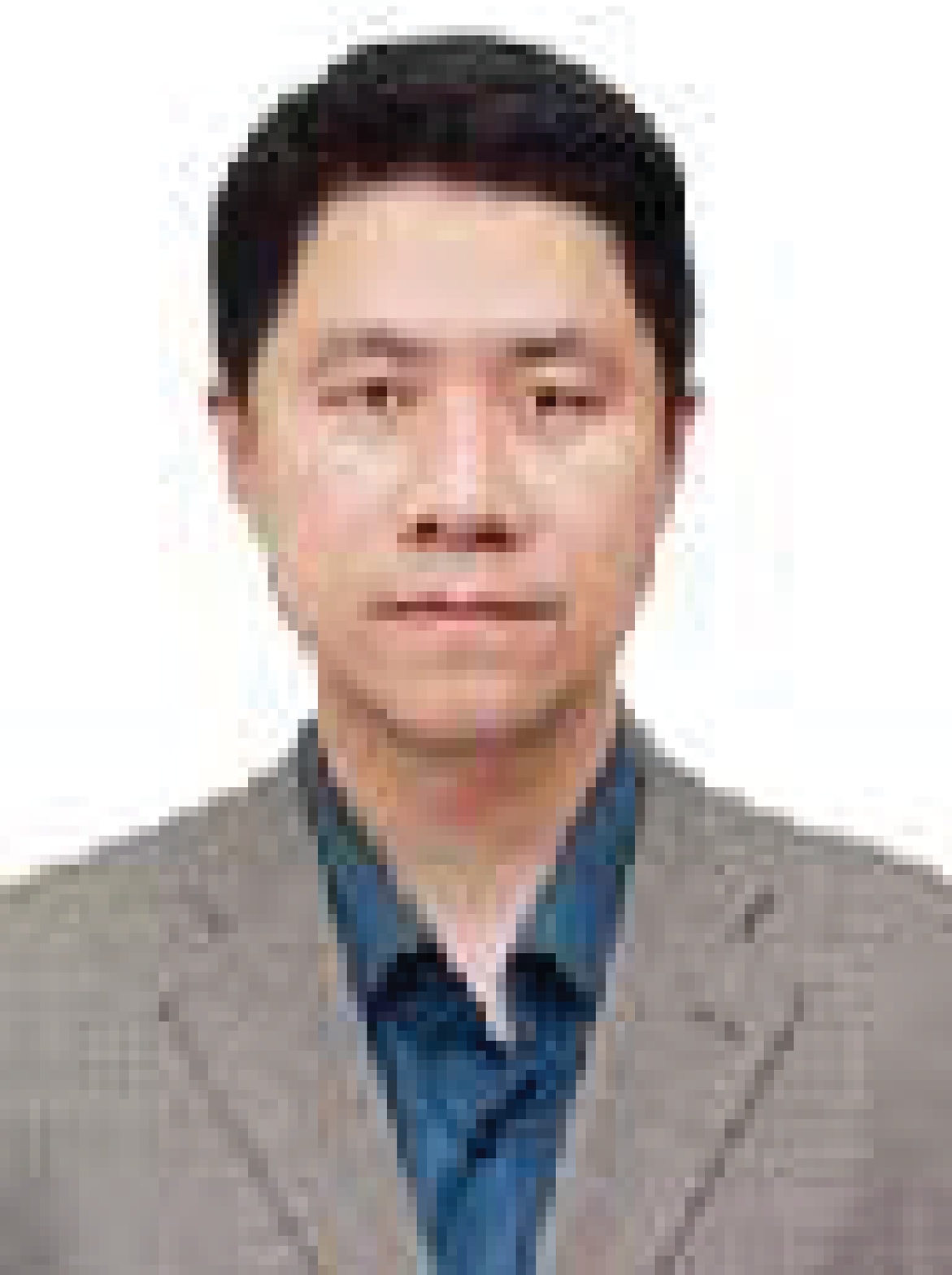}}]{Hao Jiang}
received the BS degree and the Ph.D. degree in communication networks from Wuhan University, Wuhan, China, in 1998 and 2004, respectively.
He is currently a Professor at the School of Electronic Information, Wuhan University. His research interests include wireless LANs, wireless ad hoc network, and vehicle ad hoc network.
\end{IEEEbiography}

\begin{IEEEbiography}[{\includegraphics[width=1in,height=1.25in,clip,keepaspectratio]{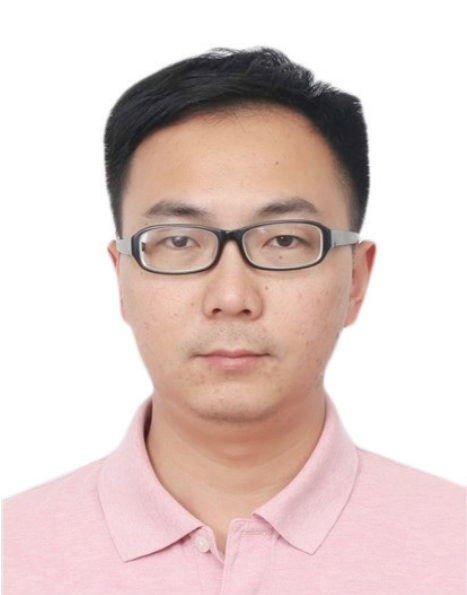}}]{Shimin Gong}
received the B.E. degree in electrical engineering from Huazhong University of Science and Technology, Wuhan, China, in 2009, and the Ph.D. degree in computer engineering from Nanyang Technological University, Singapore, in 2014. He is currently an Associate Professor with the Shenzhen Institutes of Advanced Technology, Chinese Academy of Sciences. He was a visiting scholar at The Chinese University of Hong Kong, Shatin, Hong Kong in 2011, and the University of Waterloo, Waterloo, ON, Canada, in 2012. His research interests include wireless powered
D2D networks, mobile edge computing, backscatter communications and networking.
\end{IEEEbiography}




\end{document}